\documentclass[%
 reprint,
superscriptaddress,
 amsmath,amssymb,
 aps,
 pra,
floatfix,
]{revtex4-2}

\usepackage{graphicx}
\usepackage{microtype}
\usepackage{dcolumn}
\usepackage{bm}
\usepackage{hyperref}
\hypersetup{hidelinks}
\usepackage{tikz}
\usepackage{booktabs}
\usepackage{capt-of}
\usetikzlibrary{shapes, arrows.meta, positioning, optics} 

\begin{document}
\providecommand{\NTwoSQRho}{-0.098747}
\providecommand{\NTwoSQGainPP}{0.417}
\providecommand{\PaperNTwoRhoPointCount}{401}
\providecommand{\NTwoDQRho}{-0.081547}
\providecommand{\NTwoDQGainPP}{3.685}
\providecommand{\NThreeSQRho}{-0.041677}
\providecommand{\NThreeSQGainPP}{0.412}
\providecommand{\PaperNThreeRhoPointCount}{263}
\providecommand{\NThreeDQRho}{-0.024515}
\providecommand{\NThreeDQGainPP}{3.669}
\providecommand{\PaperSQBenchmarkR}{0.25}
\providecommand{\PaperDQBenchmarkR}{0.50}

\title{Squeezing-preserving coherent control in Fourier-shaped optical lattices}

\author{Ko-Wei Cheng}
\email{b11202005@ntu.edu.tw}
\author{Wui Seng Leong}
\affiliation{Department of Physics, National Taiwan University, Taipei 10617, Taiwan}
\author{Shau-Yu Lan}
\affiliation{Department of Physics, National Taiwan University, Taipei 10617, Taiwan}
\affiliation{Institute for Atomic and Molecular Sciences, Academia Sinica, Taipei 10617, Taiwan}
\affiliation{Center for Quantum Science and Engineering, National Taiwan University, Taipei 10617, Taiwan}

\date{October 31, 2026}

\begin{abstract}
Motional squeezed states of atoms in optical lattices are affected by the anharmonicity of the trapping potential. Unequal vibrational energy spacings cause dephasing among the occupied motional-state coherences, leading to distortion of the wavepacket and a reduction in squeezing. We show numerically that adding a weak second or third spatial harmonic to a one-dimensional lattice slows the initial decay of the squeezing. Squeezed states are prepared by one or two sudden changes of the lattice depth. For each harmonic and each preparation scheme, we choose the harmonic amplitude that keeps the evolved state closest, on average over 1~ms, to its evolution in a harmonic trap. We then compare the squeezing with that in a sinusoidal lattice with the same initial squeezing and trap frequency. In all four cases studied, the successive squeezing maxima initially decay more slowly in the shaped lattice. The improvement arises because the optimal harmonic nearly equalizes the two dominant vibrational level spacings over the region occupied by the squeezed wave packet, which canceling the quartic term of the potential alone does not achieve. Because the amplitude of the added harmonic is static and can be produced with existing Fourier-synthesis techniques, lattice shaping offers a simple way to preserve motional squeezing without additional time-dependent control.
\end{abstract}

\maketitle

\begin{figure*}[!t]
 \centering
 \begin{tikzpicture}[
  x=1cm,y=1cm,>=Latex,font=\small,
  axis/.style={draw=black!55,densely dashed,-{Latex[length=2.2mm]}},
  direct beam/.style={draw=orange!80!black,line width=4.2pt,opacity=0.28,line cap=round},
  projected beam/.style={draw=orange!80!black,line width=4.2pt,opacity=0.28,line cap=round},
  fundamental beam/.style={draw=blue!75!black,line width=4.2pt,opacity=0.28,line cap=round},
  harmonic beam/.style={draw=orange!80!black,line width=4.2pt,opacity=0.28,line cap=round},
  direct ray/.style={draw=orange!85!black,line width=0.9pt,-{Latex[length=2.2mm]}},
  projected ray/.style={draw=orange!85!black,line width=0.9pt,-{Latex[length=2.2mm]}},
  fundamental ray/.style={draw=blue!80!black,line width=0.9pt,-{Latex[length=2.2mm]}},
  harmonic ray/.style={draw=orange!85!black,line width=0.9pt,-{Latex[length=2.2mm]}},
  atom/.style={circle,draw=black!75,fill=green!25,minimum size=7pt,inner sep=0pt},
  beam label/.style={font=\footnotesize,fill=white,inner sep=1.5pt}
]
  \node[font=\bfseries] at (-4.45,1.72) {(a) single-wavelength scheme};
  \draw[axis] (-8.35,0) -- (-0.55,0);
  \node[font=\footnotesize] at (-0.68,0.34) {$x$};

  \draw[direct beam] (-8.00,0.12) -- (-0.90,0.12);
  \draw[direct beam] (-0.90,-0.12) -- (-8.00,-0.12);
  \draw[direct ray] (-7.90,0.12) -- (-4.98,0.12);
  \draw[direct ray] (-1.00,-0.12) -- (-3.92,-0.12);
  \node[beam label] at (-6.75,0.36) {axial pair};

  \draw[projected beam] (-5.36,-1.35) -- (-3.54,1.35);
  \draw[projected beam] (-3.54,-1.35) -- (-5.36,1.35);
  \draw[projected ray] (-5.31,-1.27) -- (-4.75,-0.45);
  \draw[projected ray] (-3.59,-1.27) -- (-4.15,-0.45);
  \node[beam label] at (-4.45,-1.62) {angled pair};

  \draw[black!70] (-3.52,0.20) arc[start angle=12,end angle=48,radius=0.95];
  \node[font=\footnotesize] at (-3.84,0.35) {$\theta$};
  \node[atom] at (-4.45,0) {};

  \draw[black!18] (0,-1.82) -- (0,1.82);

  \node[font=\bfseries] at (4.45,1.72) {(b) two-color scheme};
  \draw[axis] (0.55,0) -- (8.35,0);
  \node[font=\footnotesize] at (8.22,0.34) {$x$};

  \draw[fundamental beam] (0.95,0.16) -- (7.95,0.16);
  \draw[fundamental ray] (1.05,0.24) -- (3.95,0.24);
  \draw[fundamental ray] (7.85,0.08) -- (4.95,0.08);
  \node[beam label] at (2.05,0.52) {$\lambda_0$};

  \draw[harmonic beam] (0.95,-0.20) -- (7.95,-0.20);
  \draw[harmonic ray] (1.05,-0.12) -- (3.95,-0.12);
  \draw[harmonic ray] (7.85,-0.28) -- (4.95,-0.28);
  \node[beam label] at (6.65,-0.55) {$\lambda_0/n$};

  \node[atom] at (4.45,0) {};
\end{tikzpicture}
 \caption{Two schemes for a lattice with a fundamental of wave vector $k_0=2\pi/\lambda_0$ and a harmonic of wave vector $nk_0$; in both, the lattice period is $d_0=\lambda_0/2$. (a) Single-wavelength scheme. All beams have the wavelength $\lambda_0/n$. The axial pair produces the harmonic, and the angled pair, with each beam at $\theta=\cos^{-1}(1/n)$ to the lattice axis, produces the fundamental. (b) Two-color scheme with standing waves at $\lambda_0$ and $\lambda_0/n$.}
 \label{fig:lattice_setup}
\end{figure*}
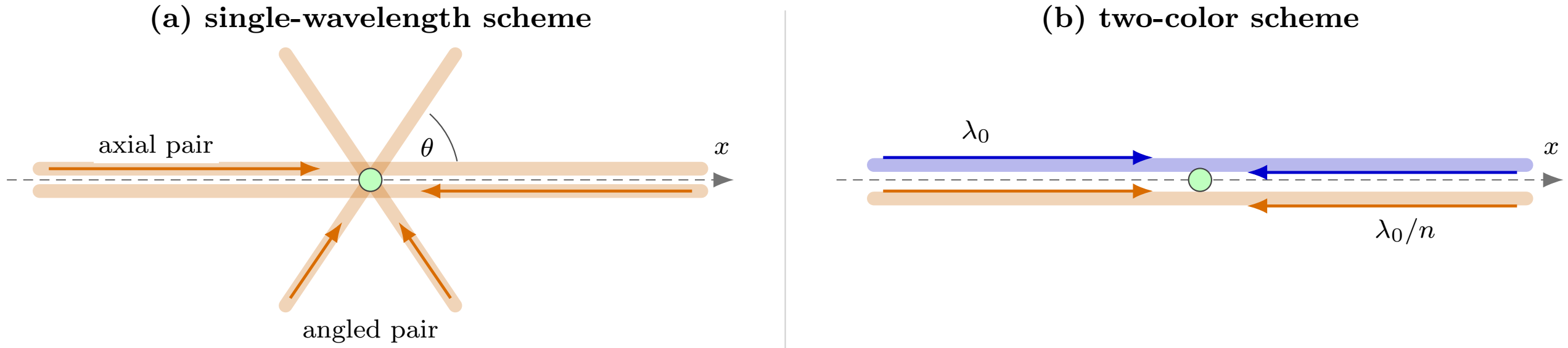

\section{Introduction}

The quantized motion of trapped atoms provides a platform for preparing and controlling nonclassical states. Motional squeezed states were demonstrated in trapped ions~\cite{meekhof1996generation}, and methods for coherent control of trapped-ion motion are reviewed in Ref.~\cite{leibfried2003quantum}. Squeezed motional states were also prepared in optical lattices with neutral atoms~\cite{morinaga1999manipulation}. Sudden changes in trapping frequency provide a rapid way to generate single-mode squeezing~\cite{xin2021rapid}, which has also been used in the preparation of two-mode squeezed states in atomic mechanical oscillators~\cite{leong2023creation}.

A squeezed wavepacket evolves periodically in a harmonic potential because the vibrational energy levels are equally spaced. In an optical lattice, however, the potential is anharmonic. A quench populates several vibrational levels, and their unequal energy spacings cause the components of the wavepacket to accumulate different relative phases. The resulting dephasing distorts the otherwise periodic squeezing dynamics. Anharmonic effects have also been studied in mechanical oscillators~\cite{marti2024quantum}, optical tweezers~\cite{beguin2025generation}, and frequency-switching protocols for trapped ions~\cite{alonso2013quantum}. Rephasing of breathing motion by additional excitation has been explored in optical lattices~\cite{bulatov1998echo}.

Here we examine a complementary approach: modifying the lattice potential itself by adding a second or third spatial harmonic. Fourier synthesis of optical potentials has been demonstrated experimentally and used to control lattice geometry and atomic transport~\cite{Ritt2006,salger2007atomic,kosch2022multifrequency}. The added harmonic provides a way to modify the vibrational spectrum without changing the basic quench protocol. The physical question is whether an appropriate lattice shape can make the transitions populated by a squeezed wavepacket more nearly equal in energy and thereby reduce distortion during its evolution.

We first identify lattice shapes that maximize the time-averaged fidelity between the full lattice evolution and its harmonic-oscillator reference, an ideal oscillator whose frequency at each stage is set by the gap between the two lowest even levels of that well. We then examine how these shapes change the occupied vibrational spectrum and compare their squeezing dynamics with a sinusoidal lattice. We call the lattice with the added harmonic the shaped lattice and the sinusoidal lattice the unshaped control. For the latter comparison, the initial squeezing and effective trapping frequency are matched, so the observed differences concern subsequent dynamics rather than the amount of squeezing initially prepared. The calculations retain the complete wavefunction over many lattice wells; a separate spatial-filter calculation is used only to help interpret the influence of probability outside the initial well.

\section{Model and methods}

\subsection{Lattice with an added harmonic}

We consider a one-dimensional optical lattice along $x$ with fundamental wave vector $k_0=2\pi/\lambda_0$ and period $d_0=\lambda_0/2$, where $\lambda_0=1064$ nm. We add a standing wave with wave vector $nk_0$, where $n=2$ or 3. Figure~\ref{fig:lattice_setup} shows two ways to produce this potential, discussed in Sec.~\ref{sec:experimental}. The potential is
\begin{equation}
 V_n(x)=-\frac{V_0}{1+\rho}
 \left[\cos^2(k_0x)+\rho\cos^2(nk_0x)\right],
 \label{eq:potential}
\end{equation}
where $V_0$ is an energy scale held fixed during each stage and changed at a quench, and $\rho$ is the amplitude of the harmonic relative to the fundamental. The parameter $\rho$ can be positive or negative. The factor $1/(1+\rho)$ fixes the potential minimum at $V_n(0)=-V_0$. For $\rho=0$, Eq.~\eqref{eq:potential} reduces to the sinusoidal lattice $-V_0\cos^2(k_0x)$ for both values of $n$. We call this the unshaped lattice and use it as the common control. The barrier height between wells is
\[
 \Delta V_n=\max_{|x|\leq \pi/(2k_0)}V_n(x)-V_n(0),
\]
which generally differs from $V_0$. For simplicity, we still call $V_0$ the lattice depth.

We measure energy in units of the recoil energy $E_{R,0}=\hbar^2k_0^2/(2M)$, where $M$ is the atomic mass, and time in units of $t_{R,0}=\hbar/E_{R,0}$. For ${}^{87}\mathrm{Rb}$, $E_{R,0}/h\approx 2.03$ kHz and $t_{R,0}\approx 78.5$ $\mu\mathrm{s}$. With $\xi=k_0 x$, $\tilde V_0=V_0/E_{R,0}$, and $\tilde t=t/t_{R,0}$, the Schrödinger equation reads
\begin{equation}
 i\frac{\partial\psi}{\partial\tilde t}
 =\left[-\frac{\partial^2}{\partial\xi^2}
 -\frac{\tilde V_0}{1+\rho}\left(\cos^2\xi+\rho\cos^2(n\xi)\right)\right]\psi .
 \label{eq:dimless_se}
\end{equation}
A tilde marks a dimensionless quantity; for example, a frequency $\omega$ becomes $\tilde\omega=\omega t_{R,0}$.

Near the well minimum, the dimensionless potential is
\begin{equation}
 \tilde U(\xi;\tilde V_0)=-\tilde V_0+\alpha_2\xi^2+\alpha_4\xi^4+\mathcal O(\xi^6),
 \label{eq:taylor}
\end{equation}
with
\begin{equation}
 \alpha_2=\tilde V_0\frac{1+n^2\rho}{1+\rho},\qquad
 \alpha_4=-\frac{\tilde V_0}{3}\frac{1+n^4\rho}{1+\rho}.
\end{equation}
The coefficient $\alpha_2$ vanishes at $\rho=-1/n^2$. For more negative $\rho$, each lattice site splits into a double well, so we consider only $\rho\geq -1/n^2$. The quartic coefficient $\alpha_4$ vanishes at $\rho=-1/n^4$, i.e., at $-0.0625$ for $n=2$ and $\approx -0.0123$ for $n=3$. At this point the well is harmonic up to fourth order, but higher-order terms remain. A squeezed wave packet is broad enough for these terms to matter, so the optimal $\rho$ need not equal $-1/n^4$ (Sec.~\ref{sec:physical_origin}).

\subsection{Squeezing protocols}

We prepare squeezed states by suddenly changing the lattice depth $\tilde V_0$~\cite{xin2021rapid}, while $\rho$ is kept fixed. The atom starts in the ground state at the initial depth $\tilde V_0=\tilde V_{\rm ref}$. Unless stated otherwise, $\tilde V_{\rm ref}=500$, where lowest-band tunneling in the unshaped lattice is negligible within 1~ms~\cite{SupplementalMaterial,Bloch_2008}. We use two protocols, a single quench (SQ) and a double quench (DQ). The squeezing parameter $r$ sets the depth to which $\tilde V_0$ is lowered:
\begin{equation}
 \tilde V_{\rm s}=\tilde V_{\rm ref}e^{-4r}\quad(\mathrm{SQ}),\qquad
 \tilde V_{\rm d}=\tilde V_{\rm ref}e^{-2r}\quad(\mathrm{DQ}).
 \label{eq:protocol_depths}
\end{equation}
where $\tilde V_{\rm s}$ is the value of $\tilde V_0$ after the single quench and $\tilde V_{\rm d}$ is its intermediate value in the double quench. For an initial harmonic ground state, both protocols produce a squeezed state with squeezing parameter $r$. In the lattice, the actual squeezing differs from $r$ and from one protocol to the other.

\begin{figure*}[!t]
 \centering
 \includegraphics[width=0.92\textwidth]{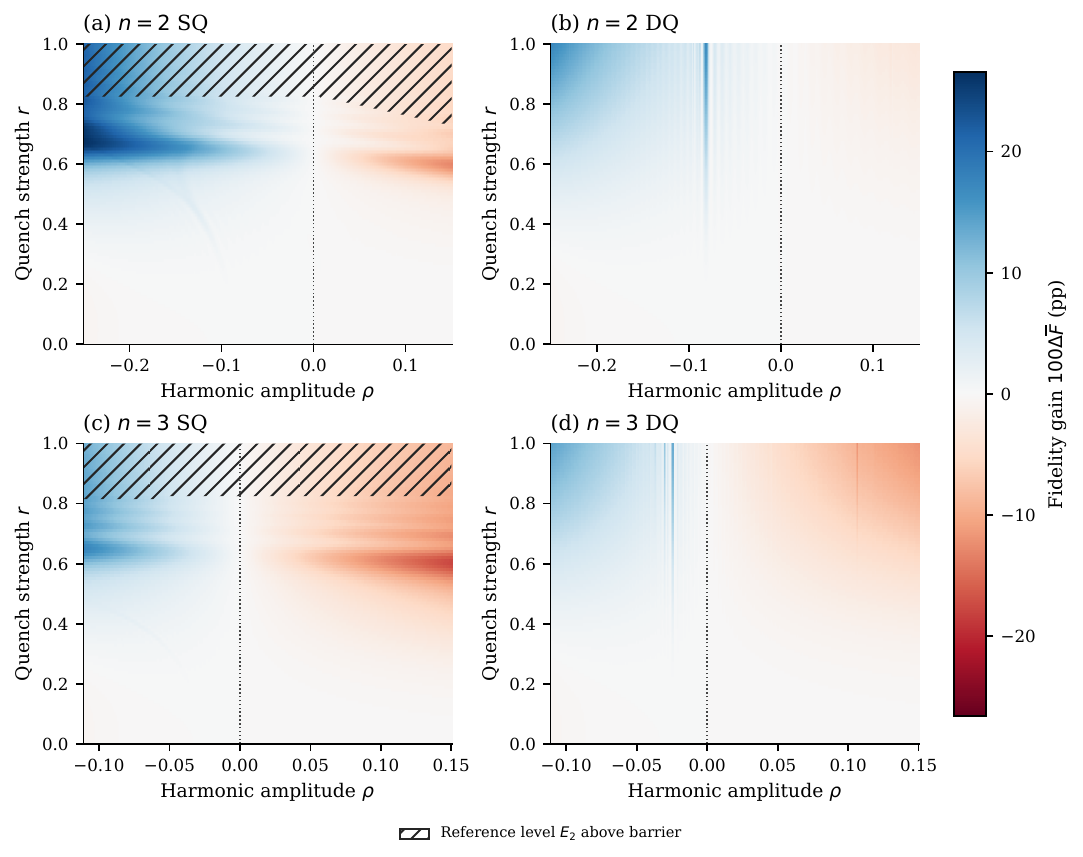}
 \caption{Fidelity gain $\Delta\bar F$ (in pp) relative to the unshaped lattice for (a) $n=2$, SQ; (b) $n=2$, DQ; (c) $n=3$, SQ; and (d) $n=3$, DQ, with $\tilde V_{\rm ref}=500$. Blue indicates a gain and red a loss. The dotted line marks $\rho=0$. Hatching marks parameters where the level $E_2$ lies above the barrier in either the shaped or unshaped lattice.}
 \label{fig:fourway_improvement}
\end{figure*}

\subsubsection{Single quench}

At $\tilde t=0$, the depth is suddenly lowered from $\tilde V_{\rm ref}$ to $\tilde V_{\rm s}$, and the state evolves in the shallower lattice. In the harmonic approximation of Eq.~\eqref{eq:taylor}, a well has the trap frequency $\tilde\omega_{\rm ho}=2\sqrt{\alpha_2}$, which scales as $\sqrt{\tilde V_0}$ for any $\rho$. The single quench therefore lowers the trap frequency by the factor $e^{2r}$, and the initial ground state becomes a squeezed state with squeezing parameter $r$~\cite{xin2021rapid}. Its position variance oscillates as
\begin{equation}
 \operatorname{Var}[\xi(\tilde t)]
 =\frac{1}{\Omega}\left[\cosh(2r)-\sinh(2r)\cos(2\Omega\tilde t)\right],
 \label{eq:var_x}
\end{equation}
where $\Omega$ is the dimensionless frequency at $\tilde V_{\rm s}$. In the lattice, the unequal level spacings distort this periodic oscillation.

\subsubsection{Double quench}

In the DQ protocol, the depth is lowered from $\tilde V_{\rm ref}$ to $\tilde V_{\rm d}$, held for a time $\tilde\tau$, and raised back to $\tilde V_{\rm ref}$:

\[
 (\tilde V_{\rm ref},\rho)\xrightarrow{\rm quench}
 (\tilde V_{\rm d},\rho)\xrightarrow{\text{hold }\tilde\tau}
 (\tilde V_{\rm d},\rho)\xrightarrow{\rm quench}(\tilde V_{\rm ref},\rho).
\]

In the harmonic approximation, the trap frequency during the hold is lower by the factor $e^r$. We set $\tilde\tau$ to a quarter of the oscillation period at the intermediate depth. Because a quench from the ground state populates only even levels, we define the oscillation frequency of a well at depth $\tilde V_\sigma$ ($\sigma$ = ref, s, or d) from its two lowest even levels:
\begin{equation}
 \tilde\omega_\sigma=\frac{\tilde E_2^{(\sigma)}-\tilde E_0^{(\sigma)}}{2},
 \qquad
 \tilde\tau=\frac{\pi}{2\tilde\omega_{\rm d}} ,
 \label{eq:dq_hold}
\end{equation}
where $\tilde E_\ell^{(\sigma)}$ is the energy of level $\ell$ at that depth, in recoil units, calculated in one well with the wave function set to zero at the cell edges. The $0\to 2$ component of the breathing motion then oscillates at $2\tilde\omega_\sigma$, as in Eq.~\eqref{eq:var_x}. In a harmonic well, all levels are equally spaced, and $\tilde\omega_\sigma$ equals the frequency $2\sqrt{\alpha_2}$ set by the curvature of the potential. In the lattice, the level spacings change with energy, so $\tilde\omega_\sigma$ differs from $2\sqrt{\alpha_2}$ and serves as an effective oscillation frequency. The hold time is thus fixed by the spectrum of the intermediate well. In physical units, $\tau=\pi\hbar/(E_2^{(\mathrm d)}-E_0^{(\mathrm d)})$. We calculate it separately for each lattice. Each lattice and its harmonic reference use the same hold.

\subsection{Numerical simulation and fidelity}

We solve Eq.~\eqref{eq:dimless_se} on a uniform grid with periodic boundary conditions spanning many lattice sites. The initial state is the ground state calculated on one periodic cell at $\tilde V_{\rm ref}$, placed in the central site; all other sites are empty. The wave function is normalized over the whole grid, including any population that tunnels out of the initial well. Because the depth is constant between quenches, the evolution within each stage is given exactly by $e^{-i\tilde H\tilde t}$, which we compute by diagonalizing $\tilde H$ in the plane-wave basis of the grid. We increased the number of sites and grid points until the results converged, and we checked the evolution against the second-order split-operator method~\cite{strang1968construction}.

To measure the distortion caused by anharmonicity, we compare the lattice state $\lvert\psi(t)\rangle$ with a reference state $\lvert\psi_{\rm id}(t)\rangle$ that evolves under the same protocol in a harmonic potential. During each stage $\sigma$, the reference evolves under
\begin{equation}
 \tilde H_\sigma^{\rm id}
 =-\frac{\partial^2}{\partial\xi^2}+\frac{\tilde\omega_\sigma^2}{4}\,\xi^2 ,
 \label{eq:harmonic_reference}
\end{equation}
whose frequency $\tilde\omega_\sigma$ is set by that lattice well at the same depth [Eq.~\eqref{eq:dq_hold}]. We use this level-spacing frequency rather than the curvature frequency $2\sqrt{\alpha_2}$, because the reference must reproduce the lowest even-parity transition of that well; a mismatch in frequency alone would otherwise reduce the fidelity. The quenches occur at the same times as in the lattice. The reference starts from the same state as the lattice simulation, the lattice ground state at $\tilde V_{\rm ref}$, projected onto a converged harmonic-oscillator basis on the real line~\cite{SupplementalMaterial}. The two states therefore coincide before the first quench, and their subsequent difference measures the departure from harmonic evolution. No parameter of the reference is adjusted. The fidelity and its time average are
\begin{equation}
 F(t)=\left|\langle\psi_{\rm id}(t)|\psi(t)\rangle\right|^2,\qquad
 \bar F=\frac{1}{N_t}\sum_{j=0}^{N_t-1}F(t_j),
 \label{eq:trajectory_fidelity}
\end{equation}
where the $N_t$ times $t_j$ are equally spaced over 1~ms, starting at $t_0=0$. Time is measured from the quench (SQ) or the return quench (DQ). In DQ the states have already evolved during the hold, so $F(0)$ can be below unity. We reduced the time between samples until $\bar F$ converged. The time average $\bar F$ reduces the influence of accidental rephasing at isolated times. We use $\bar F$ to choose $\rho$.

For large $r$ in SQ, the level $E_2$ can lie above the barrier. Equation~\eqref{eq:dq_hold} then no longer describes a bound vibration, and we mark these regions in Fig.~\ref{fig:fourway_improvement}. Throughout the DQ scan and at the working points of Sec.~\ref{sec:results}, both $E_0$ and $E_2$ lie below the barrier.

\begin{figure*}[!t]
 \centering
 \includegraphics[width=0.92\textwidth]{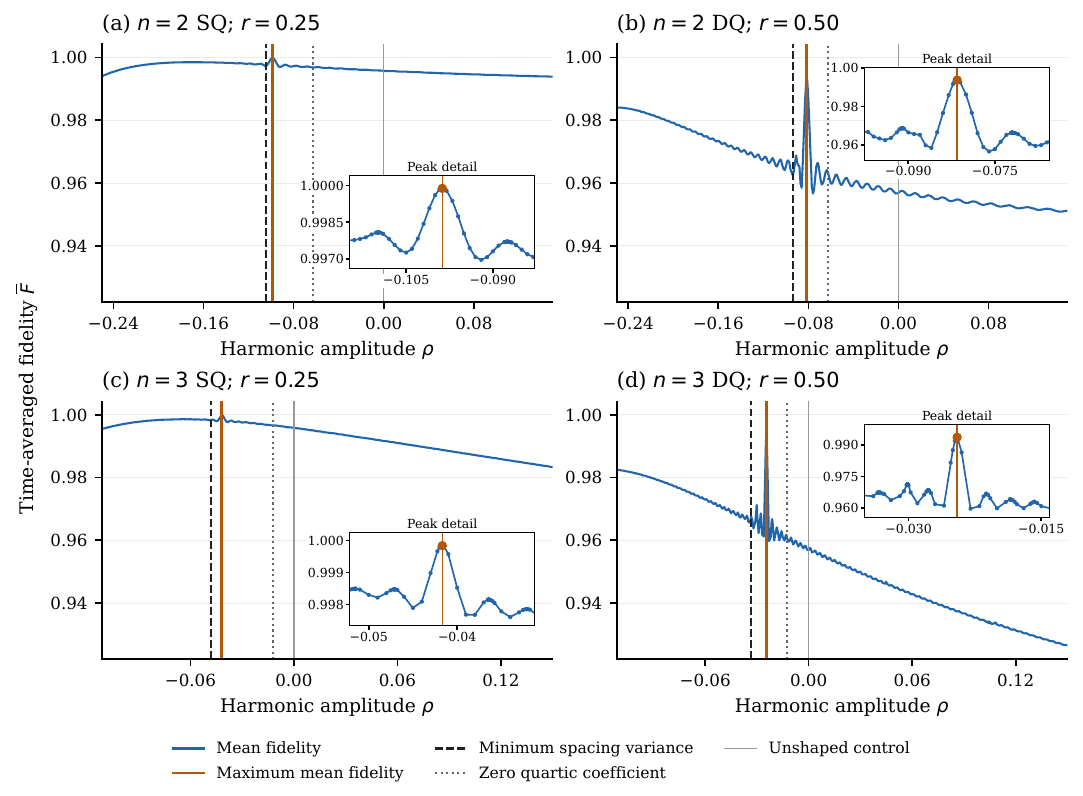}
 \caption{Time-averaged fidelity $\bar F$ versus $\rho$ at the working points for (a) $n=2$, SQ; (b) $n=2$, DQ; (c) $n=3$, SQ; and (d) $n=3$, DQ. Vertical lines mark the fidelity maximum $\rho^*$ (orange), the minimum of $\mathcal V_\Delta$ (black dashed), the zero of the quartic coefficient at $-1/n^4$ (gray dotted), and $\rho=0$ (thin gray). Insets enlarge the region near $\rho^*$.}
 \label{fig:theory_fixed_r}
\end{figure*}

\subsection{Squeezing}

The fidelity does not directly give the quadrature variances, so we also compute the squeezing. We use the full wave function, including the population in all sites. With $\delta x=x-\langle x\rangle$ and $\delta p=p-\langle p\rangle$, the covariance matrix is~\cite{weedbrook2012gaussian}
\begin{equation}
 \Gamma(t)=\begin{pmatrix}
 \langle(\delta x)^2\rangle & \frac12\langle\{\delta x,\delta p\}\rangle\\
 \frac12\langle\{\delta x,\delta p\}\rangle & \langle(\delta p)^2\rangle
 \end{pmatrix},
 \label{eq:covariance_matrix}
\end{equation}
with $\{A,B\}=AB+BA$. We scale $x$ and $p$ by the zero-point fluctuations of the final trap,
\begin{equation}
 X_{\rm zpf}=\sqrt{\frac{\hbar}{2M\omega_{\rm f}}},\qquad
 P_{\rm zpf}=\sqrt{\frac{\hbar M\omega_{\rm f}}{2}},
\end{equation}
where $\omega_{\rm f}=\tilde\omega_{\rm f}/t_{R,0}$ is the trap frequency of the final lattice from Eq.~\eqref{eq:dq_hold}, with $\mathrm f=\mathrm s$ for SQ and $\mathrm f=\mathrm{ref}$ for DQ. With $D=\operatorname{diag}(X_{\rm zpf},P_{\rm zpf})$, the quadrature at angle $\vartheta$ is
\[
 q_\vartheta=\frac{\delta x}{X_{\rm zpf}}\cos\vartheta+
 \frac{\delta p}{P_{\rm zpf}}\sin\vartheta .
\]
Its minimum variance over $\vartheta$ is $\lambda_-$, the smaller eigenvalue of $D^{-1}\Gamma D^{-1}$. Both eigenvalues equal 1 for the harmonic ground state. We define the squeezing along the principal axis of the uncertainty ellipse as
\begin{equation}
 S_-(t)=-10\log_{10}\lambda_-(t).
 \label{eq:principal_squeezing}
\end{equation}
A positive $S_-$ means a variance below the vacuum level along some quadrature. $S_-$ does not depend on the orientation of the ellipse and does not assume a Gaussian state.

Because $S_-$ is computed from the full wave function, it is sensitive to population far from the initial well. A fraction $\epsilon$ of the probability at a distance $L$ adds about $\epsilon L^2$ to $\langle(\delta x)^2\rangle$. Its effect on $S_-$ also depends on the momentum variance and the position--momentum covariance. Population that tunnels into distant wells can contribute to the deep dips in $S_-(t)$. A separate spatial-filter calculation tests this interpretation~\cite{SupplementalMaterial}. We therefore use the successive maxima of $S_-(t)$, which we call the upper envelope, to follow the squeezing within the initial well.

\subsection{Single-well spectral analysis}
\label{sec:reduced_theory}

To interpret the results, we analyze the spectrum of a single well, $|\xi|<\pi/2$, with the wave function set to zero at the cell edges. The breathing of the wave packet comes mainly from coherences between levels $m$ and $m+2$. These oscillate at angular frequencies $\Delta_m/\hbar$, where $\Delta_m=E_{m+2}-E_m$, with amplitudes $A_m=c_m^*c_{m+2}\langle m|\xi^2|m+2\rangle$. Here $c_m$ are the amplitudes of the prepared state in the eigenstates $\lvert m\rangle$ of the final well at the start of observation. In DQ they include the phases accumulated during the hold. In a harmonic well, all $\Delta_m$ are equal and the coherences stay in phase. To measure how unequal the relevant spacings are, we use the weighted relative variance

\begin{equation}
 \mathcal V_\Delta=
 \frac{\sum_m w_m(\Delta_m-\bar\Delta)^2}
      {\bar\Delta^2\sum_m w_m},
 \label{eq:spacing_objective}
\end{equation}
with $w_m=|A_m|$ and $\bar\Delta=\sum_m w_m\Delta_m/\sum_m w_m$. A smaller $\mathcal V_\Delta$ means more nearly equal spacings. We use $\mathcal V_\Delta$ only to interpret the results. The $\rho$ that maximizes $\bar F$ [Eq.~\eqref{eq:trajectory_fidelity}] differs from the $\rho$ that minimizes $\mathcal V_\Delta$ (Sec.~\ref{sec:physical_origin}).

\subsection{Parameter ranges}

We scan $r$ from 0 to 1 in steps of 0.01. We scan $\rho$ from $-1/4$ to 0.15 for $n=2$ (\PaperNTwoRhoPointCount{} values) and from $-1/9$ to 0.15 for $n=3$ (\PaperNThreeRhoPointCount{} values). Each range includes the quartic minimum at $\rho=-1/n^2$.

All scans use $\tilde V_{\rm ref}=500$ and an evolution time of 1~ms ($\tilde t\approx12.7$ for ${}^{87}\mathrm{Rb}$).

\begin{figure*}[!t]
 \centering
 \includegraphics[width=0.92\textwidth]{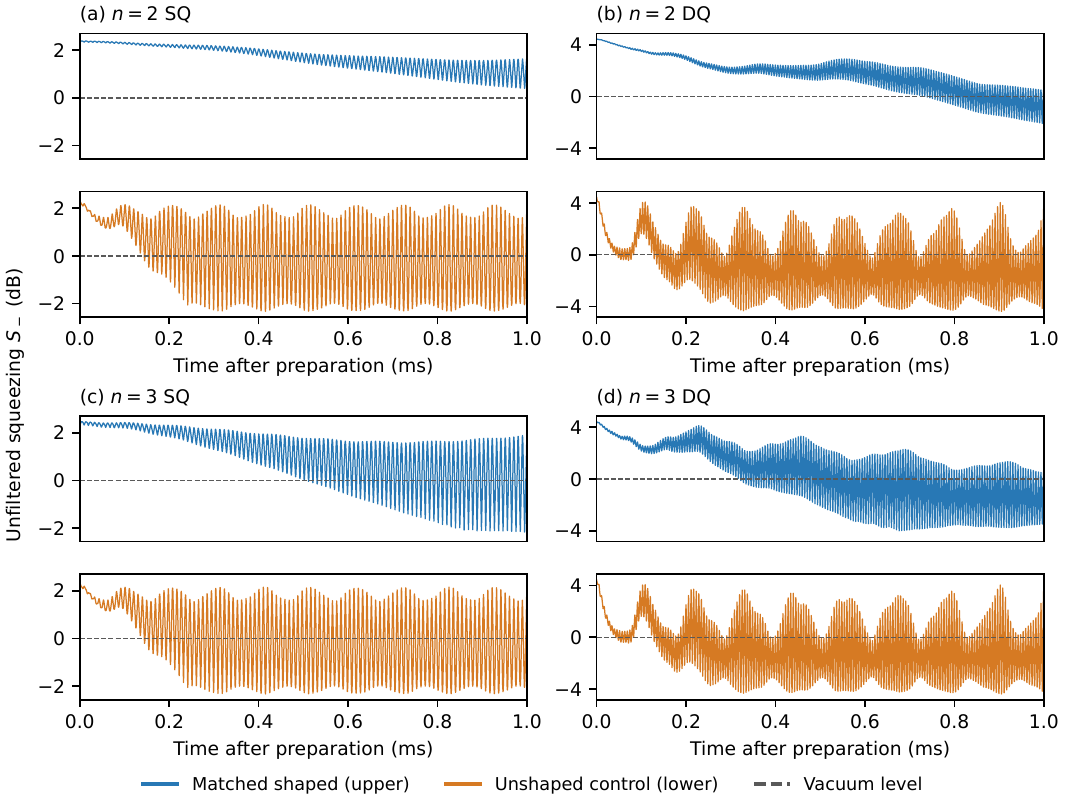}
 \caption{Squeezing $S_-(t)$ at equal initial squeezing and final trap frequency for (a) $n=2$, SQ; (b) $n=2$, DQ; (c) $n=3$, SQ; and (d) $n=3$, DQ. In each panel the upper blue trace shows the optimized lattice ($\rho=\rho^*$) and the lower orange trace shows the unshaped lattice on the same scales. The dashed gray line marks the vacuum level. Time is measured from the quench (SQ) or the return quench (DQ).}
 \label{fig:squeezing_validation}
\end{figure*}

\section{Results}
\label{sec:results}

\subsection{Choice of the harmonic amplitude}
\label{sec:fourway_comparison}

For each harmonic and protocol, we choose $\rho$ by maximizing $\bar F$ at a fixed squeezing parameter, $r=\PaperSQBenchmarkR$ for SQ and $r=\PaperDQBenchmarkR$ for DQ, which we call the working points. To isolate the effect of the harmonic, we compare each lattice with the unshaped lattice under the same protocol and the same $r$:
\begin{equation}
 \Delta\bar F(\rho,r)\equiv\bar F(\rho,r)-\bar F(0,r),
 \label{eq:delta_f}
\end{equation}
which we multiply by 100 to give the gain in percentage points (pp).

Figure~\ref{fig:fourway_improvement} shows $\Delta\bar F$ for both harmonics and both protocols. At the working points, the largest gains in all four cases occur at negative $\rho$. Elsewhere, both the size and the sign of the gain depend on $\rho$ and $r$. The maps also contain bands and narrow features. These can arise because the quenched state occupies several unequally spaced levels, whose phases add up differently in the time average as $\rho$ or $r$ changes~\cite{bakman2017collapse}.

Convergence tests support the broad gain regions; we refine the amplitude scan around each maximum~\cite{SupplementalMaterial}.

At the working points, the optimal amplitudes are $\rho^*=\NTwoSQRho$ ($n=2$) and $\NThreeSQRho$ ($n=3$) for SQ, and $\NTwoDQRho$ ($n=2$) and $\NThreeDQRho$ ($n=3$) for DQ. They raise $\bar F$ by \NTwoSQGainPP{} and \NThreeSQGainPP{} pp for SQ and \NTwoDQGainPP{} and \NThreeDQGainPP{} pp for DQ, respectively. All four values lie between the double-well limit $-1/n^2$ and the amplitude $-1/n^4$ at which the quartic coefficient vanishes.

Table~\ref{tab:matched_comparison} gives the mean fidelities in a second comparison, in which we match the initial squeezing and the final trap frequency.

\begin{table}[!ht]
 \caption{Mean fidelity over 1~ms after matching the initial squeezing and final trap frequency. Each lattice is compared with its own harmonic reference. Gains are relative to the unshaped control for the same protocol, in percentage points.}
 \label{tab:matched_comparison}
 \centering
 \begin{tabular}{lcrr}
 \toprule
 Preparation & Protocol & $\bar F$ & Gain (pp)\\
 \midrule
 $n=2$ shaped & SQ & 0.996077 & 0.036 \\
$n=3$ shaped & SQ & 0.995871 & 0.015 \\
Unshaped & SQ & 0.995719 & --- \\
$n=2$ shaped & DQ & 0.9754 & 1.85 \\
$n=3$ shaped & DQ & 0.9717 & 1.48 \\
Unshaped & DQ & 0.9569 & ---
\\
 \bottomrule
 \end{tabular}
\end{table}

\subsection{Origin of the optimal amplitude}
\label{sec:physical_origin}

Figure~\ref{fig:theory_fixed_r} shows $\bar F$ versus $\rho$ at the working points. The fidelity maximum $\rho^*$ lies away from the amplitude at which the quartic coefficient vanishes. In each panel it lies between this amplitude and the amplitude that minimizes $\mathcal V_\Delta$, so minimizing the level-spacing variance does not reproduce $\rho^*$ quantitatively. This is expected, because $\mathcal V_\Delta$ ignores the phases of the coherences, coherences between more distant levels, and tunneling into other wells.

The level spacings do, however, explain why the harmonic helps. At $\rho^*$, the $0\to 2$ and $2\to 4$ spacings differ by less than $0.005E_{R,0}$. These two coherences carry about 99.5\% (SQ) and 92\% (DQ) of the total weight $\sum_m w_m$. In the unshaped lattice and at the amplitude that cancels the quartic term, the same spacings differ by about $1\text{--}5E_{R,0}$; for these DQ comparisons, amplitude near the cell edges limits the single-well interpretation~\cite{SupplementalMaterial}. The optimal harmonic thus nearly equalizes the two transition energies that carry most of the breathing signal. Canceling the quartic term cannot do this, because it acts only near the bottom of the well~\cite{SupplementalMaterial}.

\subsection{Squeezing at equal initial squeezing}
\label{sec:matched_squeezing_results}

A higher fidelity does not by itself show that squeezing lasts longer. For the same $r$, the optimized and unshaped lattices produce different initial squeezing. We therefore adjust $\tilde V_{\rm ref}$ and $r$ for each optimized lattice, keeping $\rho=\rho^*$. The adjustment makes the initial squeezing and the final trap frequency $\omega_{\rm f}$ equal to those of the unshaped lattice at the working point. The initial squeezing is then 2.334~dB for SQ and 4.437~dB for DQ. The required values are $r=0.2760$ ($n=2$) and 0.2898 ($n=3$) for SQ, and 0.5112 and 0.5116 for DQ~\cite{SupplementalMaterial}. The DQ holds are 4.828 $\mu\mathrm{s}$ for the unshaped lattice and 4.759 and 4.762 $\mu\mathrm{s}$ for $n=2$ and 3, respectively. The optimized lattices also retain a higher mean fidelity after this adjustment (Table~\ref{tab:matched_comparison}).

Figure~\ref{fig:squeezing_validation} shows $S_-(t)$ for the four cases. In each case, the upper envelope initially decays more slowly in the optimized lattice. Near 17 $\mu\mathrm{s}$ after the DQ return, the peaks are about 4.3 and 4.1~dB for $n=2$ and 3, respectively, compared with 2.4~dB for the unshaped lattice. The deep dips are sensitive to population in distant wells. When we suppress this population with a smooth spatial filter and renormalize the state, the dips largely disappear and the filtered curves follow the upper portions of the full-state curves~\cite{SupplementalMaterial}. At later times, the squeezing oscillates and partly revives in both lattices.

\section{Discussion}

\subsection{Physical picture}

Two effects limit the squeezing in our simulations. The first is dephasing within the initial well. The added harmonic reduces it by nearly equalizing the dominant level spacings, but higher levels remain unequally spaced. Dephasing and subsequent rephasing are consistent with the slow decay and partial revivals of the upper envelope. The second effect is tunneling into distant wells, which contributes to the deep dips.

Squeezing is also limited by an effective Kerr nonlinearity in a mechanical resonator~\cite{marti2024quantum} and by the anharmonic trapping potential in optical tweezers~\cite{beguin2025generation}. The fidelity selects $\rho$, the level-spacing variance $\mathcal V_\Delta$ helps explain the choice, and the squeezing dynamics confirm the early improvement. Lattice shaping could likewise help protect other nonclassical motional states, such as two-mode squeezed states~\cite{leong2023creation} and large motional superpositions~\cite{leong2020large}.

\subsection{Scope and limitations}

Our comparison has four main limitations. First, we match only the initial squeezing and the trap frequency. The initial states and lattice depths of the two lattices therefore can still differ~\cite{SupplementalMaterial}, and equal dimensionless depths do not imply equal laser power. Second, the advantage holds only at early times. Later, the squeezing in both lattices oscillates and partly revives, and at some times the unshaped lattice shows more squeezing. Likewise, in three of the four cases at the original quench strengths the instantaneous fidelity of the optimized lattice is briefly lower than that of the unshaped lattice~\cite{SupplementalMaterial}. Third, the spatial filter changes the state, including its momentum distribution. It is an analysis tool, not a physical means of confinement~\cite{jacobs2006straightforward}. Fourth, the 1-ms evolution time is a common basis for comparison, not a physical coherence time.

The model describes a single atom moving along $x$, starting in the ground state of one well. It does not include thermal states, finite quench and readout times, three-dimensional motion, interactions, or initially occupied neighboring wells. Predictions for a specific apparatus would need these effects, together with loss and coherence times. A deeper lattice, more harmonics, or optimized quench timing may extend the useful range of $r$.

\subsection{Experimental implementation}
\label{sec:experimental}

Figure~\ref{fig:lattice_setup} shows two ways to produce the potential of Eq.~\eqref{eq:potential}. In the two-color scheme [Fig.~\ref{fig:lattice_setup}(b)], independent standing waves at $\lambda_0$ (1064~nm) and $\lambda_0/n$ (532 or 354.7~nm) produce the fundamental and the harmonic. In the single-wavelength scheme [Fig.~\ref{fig:lattice_setup}(a)], all beams have the wavelength $\lambda_0/n$. One counterpropagating pair produces the harmonic. A second pair, with each beam at the angle $\theta=\cos^{-1}(1/n)$ to the lattice axis, forms an interference pattern with the fundamental period $\lambda_0$/2. This scheme requires suppression of interference between pairs while maintaining coherence within each pair~\cite{SupplementalMaterial}.

Let $U_{\rm p}$ and $U_{\rm sh}$ be the depths of the fundamental and harmonic standing waves. Then $|\rho|=U_{\rm sh}/U_{\rm p}$. The sign of $\rho$ is set by the relative spatial phase of the two waves, 0 for $\rho>0$ and $\pi/2$ for $\rho<0$, after accounting for the sign of the atomic polarizability at each wavelength~\cite{SupplementalMaterial}. Converting the dimensionless depths into laser powers requires the polarizabilities and beam profiles. Photon-scattering rates also matter for an experimental implementation.

\section{Conclusion}

We have shown numerically that a weak second or third spatial harmonic added to a one-dimensional optical lattice slows the initial decay of motional squeezing. We chose the harmonic amplitude by maximizing the fidelity to harmonic evolution. We then compared the squeezing with that in a sinusoidal lattice with the same initial squeezing and trap frequency. In all four cases studied, the squeezing maxima initially decayed more slowly in the shaped lattice. The improvement comes mainly from a nearly equal spacing of the two dominant vibrational transitions, which canceling the quartic term does not provide.

Lattice shaping uses existing Fourier-synthesis techniques and keeps the harmonic amplitude fixed relative to the fundamental during the depth quenches. An experimental test will require the position and momentum variances and their covariance, for example through the proposed measurement of the Wigner function~\cite{winkelmann2022direct}, together with realistic quench times, light shifts, and three-dimensional confinement.

\begin{acknowledgments}
We thank Wei Chen for insightful discussions and feedback. This work was supported by the 2030 Cross-Generation Young Scholars Program of the National Science and Technology Council of Taiwan (Grant No. 112-2628-M-002-013-), the National Science and Technology Council of Taiwan (Grant No. 114-2813-C-002-003-M), the Yushan Fellow Program of the Ministry of Education of Taiwan (Grant No. NTU-113V1028-2), and the National Taiwan University Higher Education Sprout Project (Grant Nos. NTU-114L791001 and 113L7865).
\end{acknowledgments}

\begin{samepage}
\section*{Data Availability Statement}
The code, numerical data, and figure inputs that support the findings of this article have been prepared for deposition in Zenodo.
\par\end{samepage}

\bibliography{Bibliography/Sources}

\end{document}